# Thermally '*smart*' characteristics of nanofluids in parallel microchannel systems to mitigate hot spots in MEMS

**Lakshmi Sirisha Maganti [a, 1], Purbarun Dhar [b, 2], T Sundararajan [a, $] and Sarit K Das [a, b, $]**

[a] Department of Mechanical Engineering, Indian Institute of Technology Madras,

Chennai–600036, India

[b] Department of Mechanical Engineering, Indian Institute of Technology Ropar,

Rupnagar–140001, India

E–mail: [1] lakshmisirisha.maganti@gmail.com ; [2] purbarun@iitrpr.ac.in

[$] Corresponding author: skdas@iitrpr.ac.in ; tsundar@iitm.ac.in

## Abstract

Mitigation of 'hot spots' in MEMS employing in–situ microchannel systems requires a comprehensive picture of the maldistribution of the working fluid and uniformity of cooling within the same. In this article, detailed simulations employing parallel micro channel systems with specialized manifold-channel configurations i.e. U, I and Z have been performed. Eulerian-Lagrangian Discrete Phase Model (DPM) and Effective Property Model (EPM) with water and alumina-water nanofluid as working fluids have been employed. The distributions of the dispersed particulate phase and continuous phase have been observed to be, in general, different from the flow distribution and this has been found to be strongly dependent on the flow configuration. Accordingly, detailed discussions on the mechanisms governing such particle distribution patterns have been proposed. Particle maldistribution has been conclusively shown to be influenced by various migration and diffusive phenomena like Stokesian drag, Brownian motion, thermophoretic drift, etc. To understand the uniformity of cooling within the device, which is of importance in real time scenario, an appropriate figure of merit has been proposed. It has been observed that uniformity of cooling improved using nanofluid as working fluid as well as enhanced relative cooling in hot zones, providing evidence of the "smart" nature of such dispersions. To further quantify this smart effect, real-time mimicking hot-spot scenarios have been computationally probed with nanofluid as the coolant. A silicon-based microchip emitting non-uniform heat flux (gathered from real-time monitoring of an Intel® Core™ i7–4770 3.40 GHz quad core processor) under various processor load conditions has been studied and evidence of enhanced cooling of hot spots has been obtained from DPM analysis. The article sheds insight on to the behavior of non-homogeneous dispersions in complex flow domains and the caliber of nanofluids in cooling MEMS more uniformly and '*smarter*' than base fluids.

## 1. Introduction

Thermal management of MEMS (Micro Electro Mechanical Systems), not only restricted to the objective of heat removal but also to achieve uniform cooling throughout the device has become a challenge to thermofluidics researchers. This challenging situation is further aggravated by the ever-increasing heat flux dissipated by modern miniaturized devices. Non uniform cooling, often result of poor coolant flow management or highly active Joule heating in certain zones of the device, in general leads to thermal failure of the device. Therefore, there is a need for compact cooling systems with high heat removal capability as well as potential for uniform cooling as non-uniformity often leads to device failure via route of genesis of thermal hot spots. Tuckerman and Pease [1] proposed the pioneering concept of employing high area to volume ratio devices efficient in removing high heat flux from small areas. Simplistic parallel microchannel cooling systems, depending on the relative position of channels with respect to inlet and outlet manifolds can be classified into three general configurations, i.e. U, I and Z.

The major challenges that crop up in case of microchannel cooling systems are high pressure drop and non-uniform distribution of working fluid from the manifold to the parallel channels such that it leads to non-uniform cooling within the device. To meet such high thermal demands, the design of microchannel cooling system has undergone several alterations and modifications in the past decades. The extent of maldistribution of the working fluid in macro and mini channel systems is well understood from the available models [2-4]. On the contrary, the trend of maldistribution in case of microchannels is different and such models are futile for prediction of maldistribution of fluid in parallel microchannels [5]. The ultimate consequence of fluid maldistribution in parallel microchannel cooling systems is non-uniformity in cooling [6, 7] which often leads to formation of hot-spots. These hot-spots are generated by virtue of geometric parameters in addition to the non-uniformity in heat generation pattern often observed in such devices.

Attempts have been made by researchers to address flow maldistribution problems and possible solutions to resolve the same. Kumaraguruparan et al. [8] conducted a detailed study on maldistribution of fluid in U configuration systems and concluded that decrease in channel depth, width, number of channels and increase in channel length results in more uniform distribution. The problems to be addressed in case of microchannel flows have been reviewed by Kandlikar et al. [9]. Later on, focus shifted to optimizing the geometry of such complex flow systems to minimize maldistribution without compromising cooling capacity. Siva et al. [10] reported an optimization study and presented details of an optimal U type configuration with minimum maldistribution in a parallel microchannel cooling system. The performance of coolants have been reported to be enhanced by the use of nanofluids, which not only enhance heat transfer rates but have also been reported to show smart fluid characteristics with temperature [11]. Performance of microchannel cooling systems can be improved by using nanofluids [12] as working fluids [13-15]. Consequently, the non-homogeneous behavior of nanofluids was also reported [16, 17] and this placed a hurdle across modeling thermo hydraulics of nanofluids in small dimensional systems such as microchannels. Singh et al. [18] reported the thermal behavior of nanofluid in single microchannel and used DPM (Discrete Phase Model) model for understanding particle migration within the channel but the studies were limited to a single microchannel.

To the best of knowledge of the present authors, studies pertaining to maldistribution of fluid and nanoparticle concentration (since the fluids are non-homogeneous) within the channels have not been discussed. Most studies on the topic have either modeled nanofluids as homogeneous fluids (and thus have not shed any insight onto the smart behavior of such fluids with temperature) or have simply considered uniform distribution of nanofluids in the channels. The effect of flow geometry on thermal management has remained untouched as far as non-homogeneous fluids are concerned. Furthermore, locations of hotspot arising due to flow maldistribution with respect to the flow configuration have not been reported and it is an important parameter to be probed for obtaining proper device cooling. Furthermore, reports consider the heat generation by such devices to be uniform by nature and this is far from the reality. In real devices, hot spots erupt spontaneously at regions where the Joule heating is higher due to higher activity of the circuitry in that zone. The caliber of coolants and especially smart fluids such as nanofluids in mitigating such hot spots has also remained a topic, which has not been probed. Therefore, there is a need of an in-depth study to understand the behavior of nanofluid (non-homogeneous formulation) in case of three general flow configurations U, I and Z for spatially uniform and non-uniform heat generation from the device. The present article presents a detailed study on the maldistribution of nanofluids (both fluid and nanoparticle components) among parallel microchannel systems for U, I and Z configurations. The implications of such complex flow, particle distribution on the thermal distribution within the device, for both uniform and non-uniform heat generation (a real time Intel$^{TM}$ quad core i7 processor has been mimicked) has been extensively discussed, and mitigation efforts have been proposed.

## 2. Theoretical formulation

The present article deals with the distribution pattern of nanoparticles and nanofluid in the event of nanofluid flow through microchannel systems of variant configurations and its effect on the thermal performance of such systems. In order to understand the transport of nanoparticles as independent entities with respect to the fluid, an Eulerian–Lagrangian Discrete Phase Model (DPM) approach has been employed, similar to that reported by the same authors [17]. Additionally, an effective property model (EPM) approach has also been adapted to exclusively understand the non–homogeneous behavior of nanofluids.

### 2.1. Governing equations

The governing equations for EPM and continuous phase of DPM are the equations of mass, momentum and energy conservation. The equations are expressed as

$$\frac{\partial \rho}{\partial t} + \nabla \cdot (\rho \vec{V}) = 0 \tag{1}$$

$$\frac{\partial \rho \vec{V}}{\partial t} + \nabla \cdot (\rho \vec{V}\vec{V}) = -\nabla P + \nabla \cdot \left(\mu(\nabla \vec{V} + \nabla V^T)\right) + S_m \tag{2}$$

$$\rho C \left[\frac{\partial T}{\partial t} + \vec{V} \cdot \nabla T\right] = \nabla \cdot [k \nabla T] + S_e \tag{3}$$

Effects of viscous dissipation and compressibility have been considered negligible in the energy equation. In the above equations, $\rho$ is density of liquid, $V$ is velocity of the liquid, $t$ is time, $P$ is pressure, $g$ is the acceleration due to gravity, $C$ is the specific heat of fluid, $k$ is thermal conductivity of fluid and $T$ is fluid temperature. $S_m$ and $S_e$ are source terms representing exchange of momentum and energy between the continuous and discrete phase (nanoparticles). Considering a Lagrangian system of reference, the governing equation for the motion of the nanoparticles can be expressed based on Newton's second law as

$$\frac{dV_p}{dt} = F_D(u - u_p) + \frac{g(\rho_P - \rho)}{\rho_P} + F \tag{4}$$

$$F = F_G + F_B + F_T + F_L + F_P + F_V \tag{5}$$

$V_p$ is the instantaneous velocity of the particles and $F$ is the total specific force experienced by the particle. Terms $F_D$, $F_G$, $F_B$, $F_T$, $F_L$, $F_P$ and $F_V$ represent the forces resulting out of fluidic drag, gravity, Brownian motion, thermophoretic drift, Saffman lift, contribution of pressure gradient and contribution of virtual mass respectively. The forces can be expressed as [19]

$$F_D = \frac{18\mu}{\rho_p d_p^2} \frac{C_D Re}{24} \tag{6}$$

For submicron particles like nanoparticles, the classical Stokesian drag is modified to accommodate the non–continuum or slip effects (important for high Knudsen number systems, like flow over nanoparticles) at the particle–fluid interface and is expressed as

$$F_D = \frac{18\mu}{\rho_p d_p^2 C_c} \tag{7}$$

Where $C_c$ represents the Cunningham correction factor and it is the expressed as

$$C_c = 1 + \frac{2\lambda}{d_p}(1.257 + 0.4e^{-(1.1d_p/2\lambda)}) \tag{8}$$

$$F_G = \frac{g(\rho_p - \rho)}{\rho_p} \tag{9}$$

The random nature of Brownian motion when averaged over a long time period leads to zero net directional flux. Thus, a probability function is required to mathematically express the force. The amplitude of the Brownian force is expressed as

$$F_{B_i} = \zeta_i \sqrt{\frac{\pi S_0}{\Delta t}} \tag{10}$$

Where $\zeta_i$ is a random number and is part of a simpleGaussian distribution.The amplitudes of the Brownian force components are estimated at each step of the discrete phase calculations. The Brownian randomness is modeled as a spectral intensity $S_{n,ij}$ as [20]

$$S_{n,ij} = S_0 \delta_{ij} \tag{11}$$

$\delta_{ij}$ is the Kronecker delta function and the amplitude of the spectrum $S_0$ is expressed as

$$S_0 = \frac{216 \nu k_B T}{\pi^2 \rho d_p^5 (\frac{\rho_p}{\rho})^2 C_c} \tag{12}$$

Existence of temperature gradients within the fluid leads to differential bombardment by the fluid molecules on the particles, which leads to a direction drift of particles from heated towards cold regions.The phenomenon is termed as thermophoresis or Soret diffusion andthe force generated by the drift is expressed as

$$F_T = -D_{T,P} \frac{1}{m_p T} \frac{\partial T}{\partial x} \tag{13}$$

$D_{T,P}$ is the thermophoretic coefficient [21] expressed as

$$D_{T,P} = \frac{6\pi d_p \mu^2 C_s (k + C_t k_n)}{\rho (1 + 3 C_m k_n)(1 + 2k + 2 C_t k_n)} \tag{14}$$

Where $C_m=1.146$, $C_s=1.147$ and $C_t=2.18$ are momentum exchange, thermal slip and temperature jump coefficients respectively.

Saffman lift is generated by shear forces acting on the nanoparticlesdue to flow of the continuous phase and is expressed as

$$F_L = \frac{2k_s v^{\frac{1}{2}} \rho d_{ij}}{\rho_p d_p (d_{lk} d_{kl})^{1/4}} (V - V_p) \tag{15}$$

$k_s = 2.594$ is a constant and $d_{ij}$ is the deformation tensor for the continuous phase and governs the shear generated around the particle due to velocity difference between the particulate phase and fluid phase.

The force on the particles due to the pressure gradient within the fluid is expressed as

$$F_P = \left(\frac{\rho}{\rho_p}\right) V_p \frac{\partial V}{\partial x} \tag{16}$$

The inertia required to propel the fluid surrounding the particles gives rise toa virtual mass force expressible as

$$F_V = \frac{1}{2} \frac{\rho}{\rho_p} \frac{d}{dt} (V - V_p) \tag{17}$$

To determine the effective properties (density, specific heat, viscosity and thermal conductivity in ascending order of equation numbers) of the nanofluids considering EPM, the following expressions have been employed [22]

$$\rho_{nf} = (1 - \phi)\rho_{bf} + \phi \rho_p \tag{18}$$

$$(\rho C_p)_{nf} = (1 - \phi)(\rho C_p)_{bf} + \phi(\rho C_p)_p \tag{19}$$

$$\mu_{nf} = \mu_{bf}(1 + 10\phi) \tag{20}$$

$$k_{eff} = k_f \frac{[k_p + (n-1)k_{BF} - (n-1)\phi(k_{BF} - k_p)]}{[k_p + (n-1)k_{BF} + \phi(k_{BF} - k_p)]} \tag{21}$$

## 2.2. Computational details

Three dimensional parallel microchannel systems of three configurations (with respect to the relative positioning of the channels and manifolds) U, I and Z type has been used as a computational domain. The domain has been meshed and the equations for fluid flow, heat transfer, and particle dynamics has been solved employing ANSYS Fluent 14.5. Fig. 1 shows the three different geometrical configurations employed in the present study. These particular configurations have been reported as general parallel microchannel configuration systems. U has been reported to exhibit worst flow distribution [10] followed by I and the distribution is best in case of Z. Therefore, a detailed study on flow characteristics of water and nanofluid as well as particle maldistribution and the consequences of same on cooling performance of such devices are useful for design and optimization. The details of dimensions of geometry and working fluid are as follows: hydraulic diameter ($D_h$) of channel is 100μm, area ratio ($A_{channel}/A_{manifold}$) is 0.2, number of channels (N) is 7, aspect ratio of channel (H/W) is 0.1, working fluid is water and $Al_2O_3$–water nanofluid (both homogeneous and non-homogeneous dispersion formulations). A mesh consisting of quadrilateral elements has been used and the grid system at the inlet of the manifold is employed for injecting the nanoparticles in the DPM formulation. Uniform heat flux has been applied at the bottom. Sidewalls and the top wall has been considered as adiabatic for heat transfer cases. In order to understand the effects of non-uniform heat release by the device on the thermal management caliber of nanofluids, studies have also been performed with non-uniform heat flux boundary conditions applied at the bottom of the domain. The domain has been segregated into four equal zones and thermal loads have been provided based on real time data of an actual modern processor. The mesh has been thoroughly tested for grid independence of dependent variables (maldistribution parameter and steady state temperature) for all the configurations. It has been observed that any grid size suitable for the U configuration [17] is always suitable for an I or Z system with similar boundary conditions and this is justified, as U is the worst-case scenario. The flow and thermal parameters have been validated for the configurations with respect to reported data [10, 18] in accordance to protocol reported by the present authors [17].

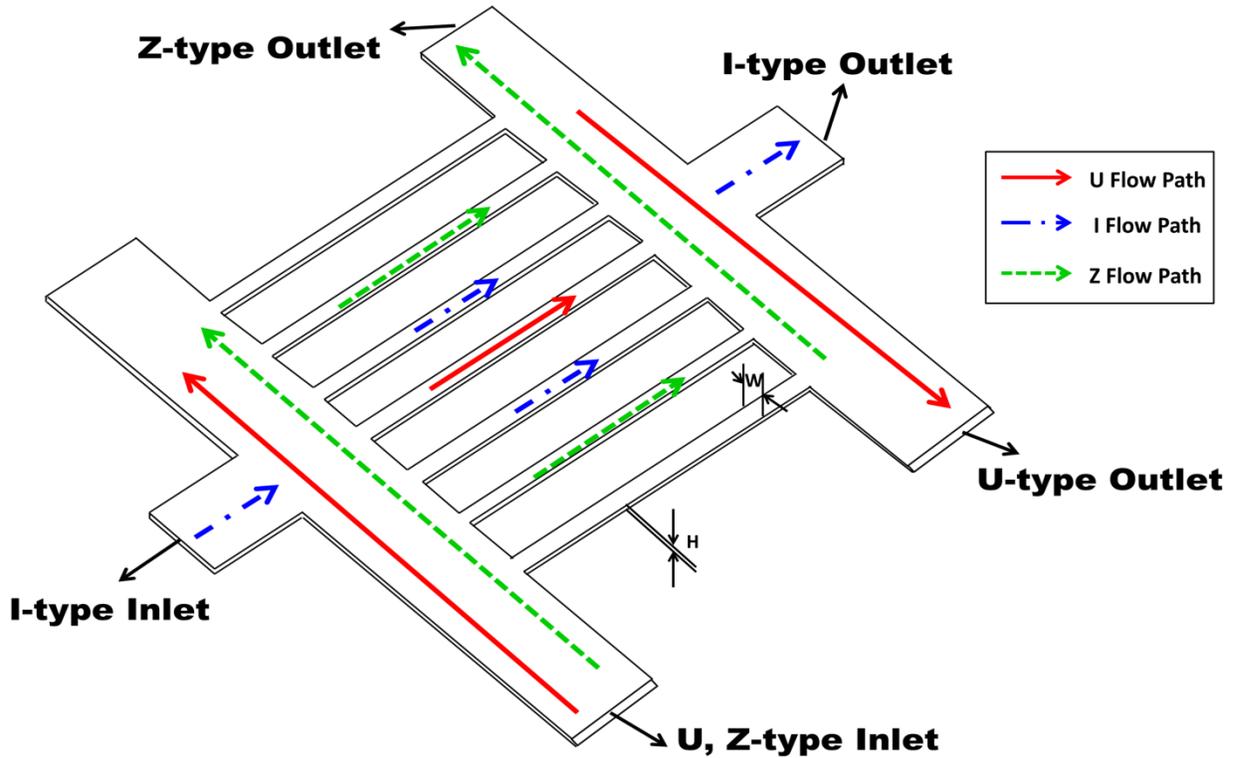

**FIGURE 1:** Schematic of the flow domains considered in the present study. U, I and Z flow configurations (manifold and channel system) have been employed and the arrows indicate the corresponding fluid distribution and flow pathways. (The complete solid domain has not been illustrated here. When one particular flow configuration is operational, the inlets and outlets of the other two types are closed. The systems have been simulated separately but are shown in the same figure for brevity).

## 3. Results and discussions

### 3.1. Adiabatic flow and particle distribution

The pressure drop in case of all three configurations (U, I and Z) provides a complete picture of the flow maldistribution of working fluid among parallel microchannels. A flow maldistribution parameter ($\eta$) and concentration maldistribution parameter ($\varepsilon$) [17] have been proposed to quantify the extent of fluid maldistribution and particle maldistribution. Channel-wise pressure experienced by nanofluid (5 vol. % alumina) and water at Re=50 has been illustrated in fig. 2 for all three configurations. From the figure, it is evident that the distribution patterns for the fluid are dissimilar for different configurations. Irrespective of configuration, the pressure drop of nanofluid is higher when compared to the water. This is expected due to higher viscosity of nanofluids. In case of U system, the magnitudes of pressure drop decrease gradually from the former channels towards the latter channels and is due to the highly non-uniform distribution of fluid in this system [17]. On the contrary, distribution of fluid is completely different for I and Z configurations. In case of I, pressure drop is high in the central channels whereas reverse situation is observed for Z configuration. The difference in fluid distribution patterns is caused primarily by the relative location of the channels with respect to the inlet and outlet manifolds. From fig. 2, it can be inferred that the distribution of fluid is more uniform in case of I and Z compared to U and hence maldistribution of working fluid is maximum for U configuration, followed by I and minimum for Z. Therefore, as far as pressure drop and consequently pumping power is involved, nanofluids do not pose too high concerns, as the viscous resistance is not very high when compared to the base fluid. Hence, they are suitable candidates as microsystem coolants.

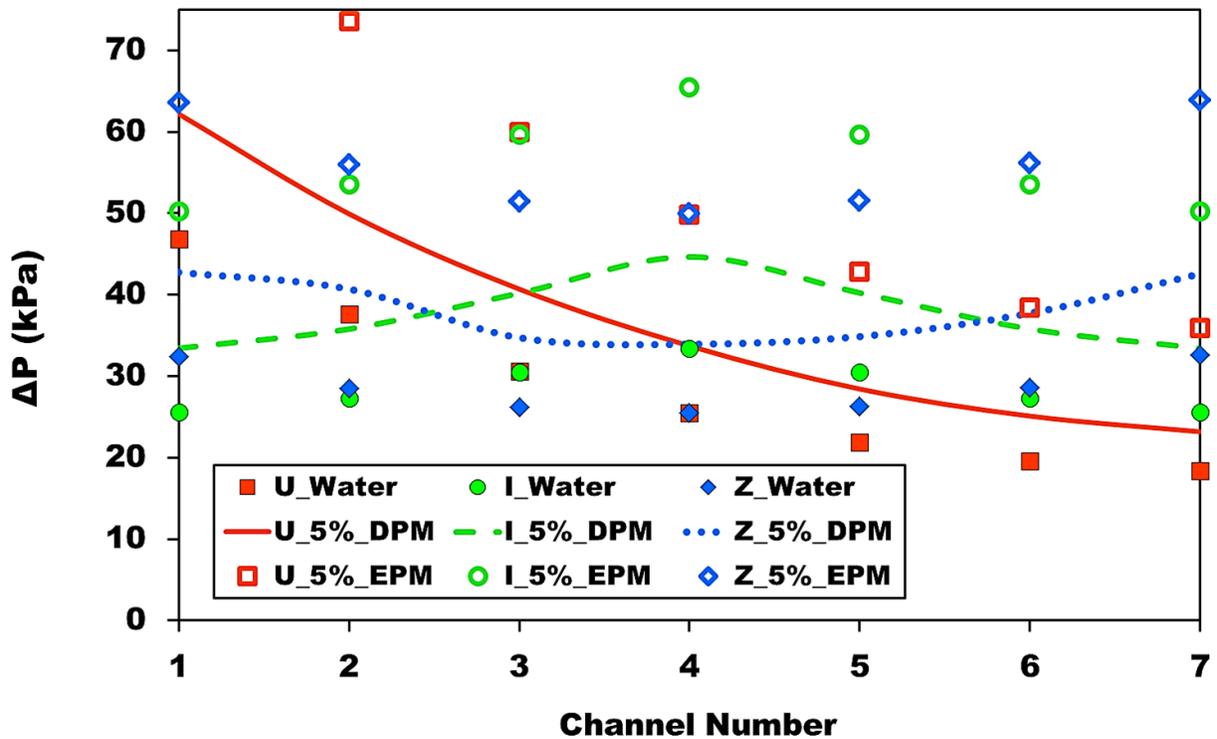

**FIGURE 2:** Pressure drop characteristics within the channels as a function of progressing channel number. Pressure drop for flow of water and 5 vol. % nanofluids (both EPM and DPM formulation) at Re 50 have been illustrated for U, I and Z flow patterns. The symmetric nature of flow distribution in I and Z configurations compared to the drastically asymmetric nature in U case can be clearly observed.

The differences observed in the trends of FMF, CMF provides insight on the non–homogeneous behavior of nanofluids in microscale flow systems, and this has been reported for a basic U configuration flow distribution system by the present authors [17]. However, the effect of flow patterns on the distribution of such non–homogeneous fluids required to be probed to estimate the thermal performance of a particular flow configuration. Considering fig. 3 (I), it can be observed that the distribution of concentration does not follow the flow distribution trends for any particular Re. At low flow inertia, increment of concentration from low to moderate involves decrease in CMF which is the result of better nanofluid distribution caused by increased localized viscosity of the fluid due to increased presence of nanoparticles. However, when the concentration is high, particle overcrowding in the inlet manifold hinders symmetric distribution of particles within the channels, leading to increment in CMF. But with increase in flow inertia from low to moderate (i.e. increasing Re), the effect on viscosity due to increment in concentration is gradually nullified and hence the CMF becomes independent of concentration. At high Re, inertia dominates over the flow regime and the viscous effects due to particles are overshadowed. Hence, the highly ordered distribution of flow ensures a similarly ordered particle distribution, thereby leading to very low CMF. In case of Z type flow pattern, the CMF is more or less same in all cases and it is due to the arrangement of channels with respect to the manifolds. The arrangement ensures highly ordered flow pattern with a high degree of symmetry and although the maldistribution of particles is high, the symmetric flow pattern ensures that the CMF is independent of the concentration.

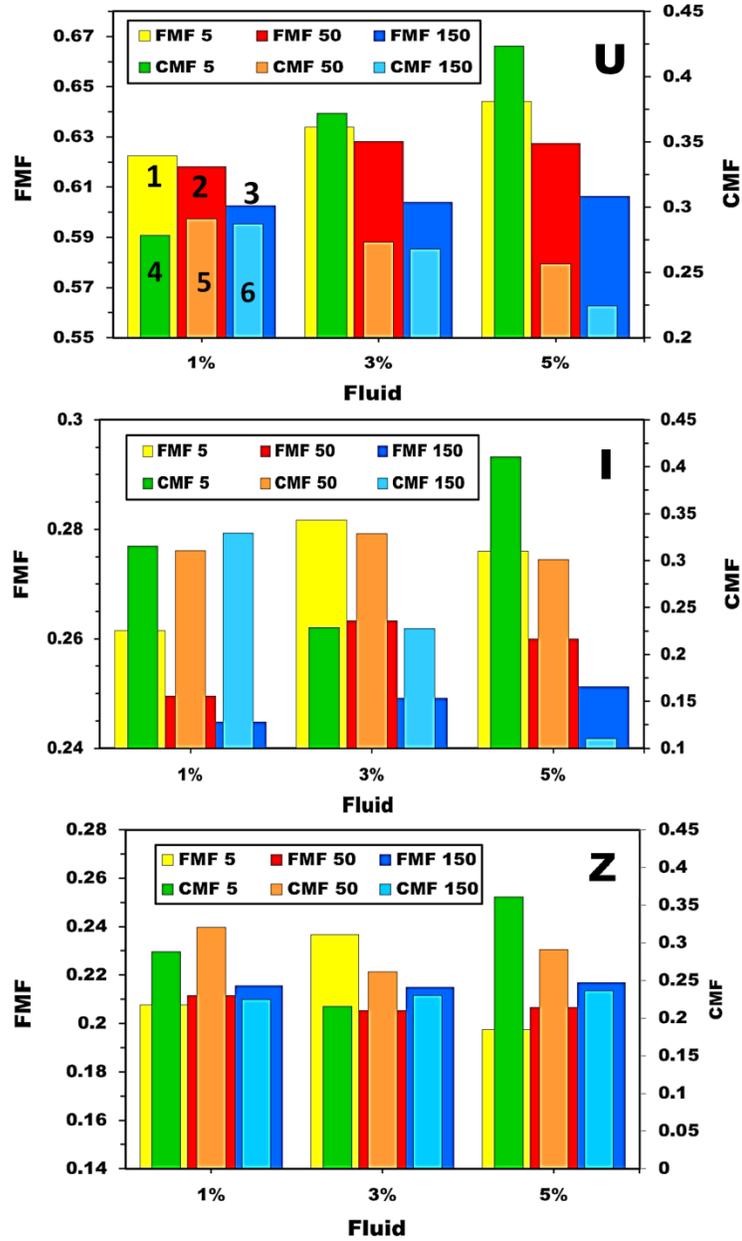

**FIGURE 3:** Comparison of FMF and CMF of U, I and Z configurations for alumina-water nanofluid at three different Reynolds numbers (5, 50 and 150) and three concentrations (1, 3 and 5 vol. %). (1, 2 and 3 represents FMF for Re 5, 50 and 150 respectively; 4, 5 and 6 represents CMF for Re 5, 50 and 150 respectively)

In particulate flows through microscale systems, the importance of particle distribution and its effects on the thermal performance of the nanofluids has been discussed [17]. It has been reported by the present authors that particle maldistribution is an important factor that governs thermal performance of the cooling system as well modulates the overall flow pattern within the system. Accordingly, it is also necessary to understand the effect of flow configuration on the effective particle concentration within each channel under steady state conditions. The effective concentration within each channel at Re = 5 and 50 for a 5 vol. % concentration nanofluid has been illustrated in Fig. 4. It can be observed that in I configuration the distribution is symmetric with respect to the central channel. As the central channel experiences highest fluid flow, the effective concentration is least as the particles are flushed away by the incoming flow to various regions within the manifold and the central channel is particle starved (but at high, viz. Re=50 and above). As the velocity within the inlet manifold decreases towards the peripheral channels, the nanoparticles are at ease to enter the channels with the fluid and hence the effective concentration increases up to the original concentration near the periphery. However, when the flow has reached the last channel, it has sacrificed a large population of nanoparticles to the preceding channels and hence the concentration in the last channel suffers a drop. However, it is noteworthy that at any instant, the average of the effective concentrations falls short of the actual concentration of the nanofluid. It can be inferred that due to the complex manifold system in I configuration, a fraction of the particle population is retained within the manifolds (especially at the far ends of the manifolds where the fluid velocity is very low and cannot flush the particles effectively into the channels). This gives rise to the impression that the average effective concentration within the channels at a point of time is lesser than the actual concentration. However, for enhanced Re, the average of the effective concentration is same as the actual concentration as the higher inertia ensures complete flushing of the nanoparticles from the manifold to the channels.

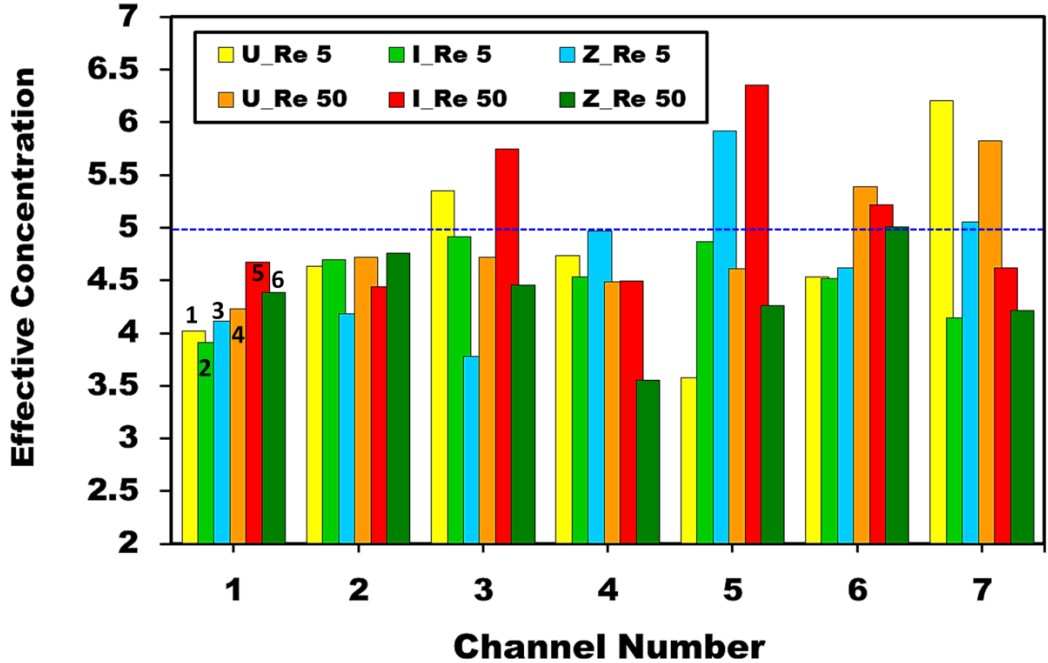

**FIGURE 4:** Channel–wise effective concentration for different flow configurations for a nanofluid (5 vol. %) flow at Re = 5 and 50. The dashed horizontal line represents the actual concentration of the nanofluid. (1, 2 and 3 represents effective concentration at Re=5 for U, I and Z respectively;4, 5 and 6 represents effective concentration at Re=50 for U, I and Z respectively)

The effective concentration distribution for a Z type system at low flow Re is similar to that of a U type system, however, with the peaks and valleys shifted due to the differences in the flow in outlet manifold and the consequent distribution from inlet manifold. However, at higher Re, the inlet manifold flushes a large population of nanoparticles towards the end regions, thereby trapping a portion of the concentration in the manifold. Similarly, the closed end of the outlet manifold, in case of higher inertia would tend to lodge a proportion of the nanoparticles due to reduced flushing at the end zone. Accordingly, the manifolds retain a certain fraction of the concentration and this leads to the effective concentration pattern similar to that of I case at low Re. Figure 5 illustrates the particle distribution contours within each channel (axially central region of channel) for different flow configurations for Re = 5. It is clearly observable from the figure (both qualitatively and quantitatively) the role of different flow domains on the distribution pattern of the particles within the channels. It has already been established from fig. 3that at up to moderate Re values, the particle distribution pattern and flow distribution pattern tend to follow different trends with respect to concentration as well as flow Re. Furthermore, the trend of CMF changes with thermal conditions as the Brownian flux enhances and the thermophoretic drift becomes active. The changes in CMF from adiabatic to diabatic systems provide insight onto the efficacy of nanofluids in such systems and the same has been illustrated in fig. 6.

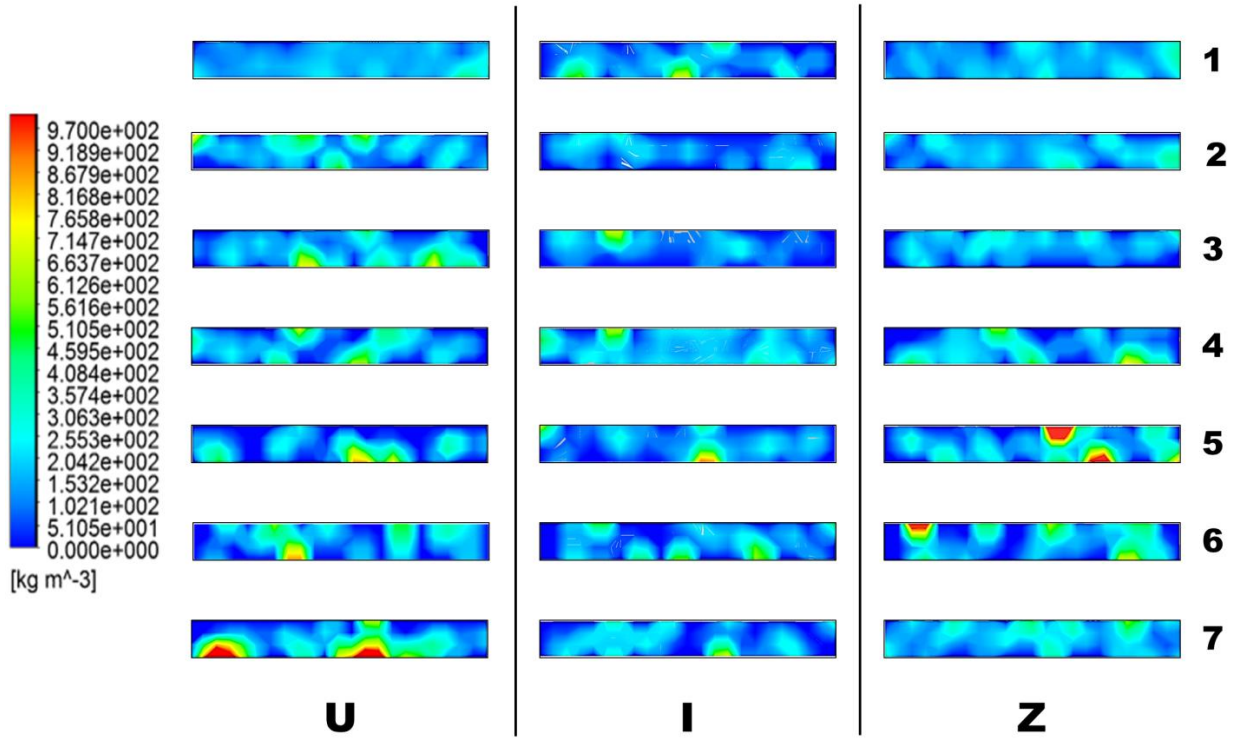

**FIGURE 5:** Channel wise (at the axially central section) particle distribution contours for different flow configurations at low Re. The effect of flow configuration on particle distribution charactersitics is clearly visible in the present figure.

### 3.2. Diabatic flow, particle and thermal distribution

The behavior of CMF with respect to heat flux and Re for U flow systems has already been reported by the present authors [17]. The behavior of CMF for I flow system has been illustrated in fig. 6 (b). At no imposed heat flux, the CMF drastically reduces with increasing Re and this is expected as the thermal slip mechanisms remain inactive due to adiabatic nature and the high inertia of the flow ensures proper distribution of the particles. As the system is subjected to medium heat flux, the CMF at higher Re is much more pronounced than the adiabatic case as the thermophoretic drift and the enhanced Brownian flux acts as an agent of maldistribution. However, as the heat flux enhances, the CMF at high Re further decreases and is similar to the adiabatic case. It is possible that the highly augmented thermophoretic drift acts in distributing the particles more evenly from the manifold to the channels over and above the action of the fluid. It is further possible that the highly reduced viscosity (as the fluid travels at elevated temperatures) causes the fluid to accelerate within the domain and flush the particles along with it. For the Z flow domain (fig. 6 (c)), the trend of CMF for adiabatic conditions and increasing Re is similar to the others and this is expected. However, the trend observed for high heat flux in I case is observed at moderate heat flux for Z case and this is probably caused by the nature of thermal distribution within the two systems. At moderate heat flux values, the Z system is somewhat symmetric in nature (with respect to cooling) like I and hence the CMF trends are similar. However, at high heat flux, maximum CMF is observed for moderate flow Re. it is probable that with high heat flux and moderate Re, the thermal profile is such that the effective thermophretic flux disrupts proper distribution of particles from the manifolds. As the Re increases, the cooling enhances and this coupled with reduction in viscosity leads to better fluid accelaration based transport and distribution of nanoparticles.

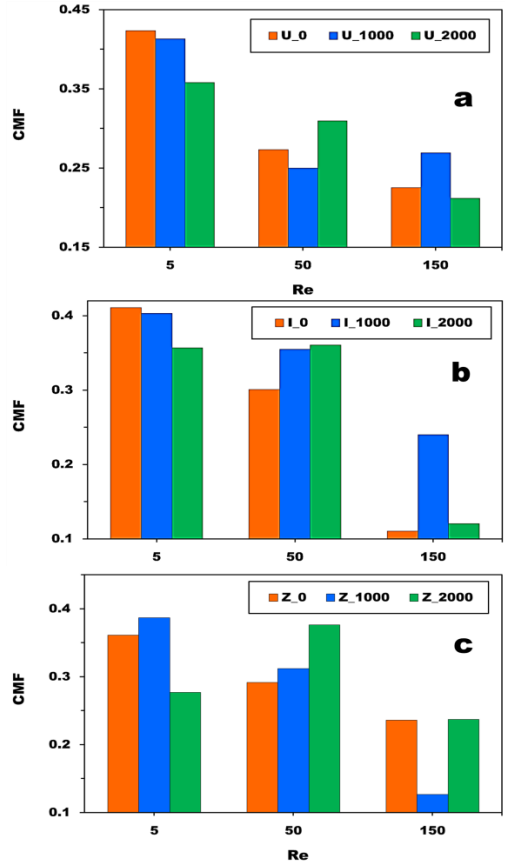

**FIGURE 6:** Illustration of the CMFs for various configurations and flow Re at different heat fluxes for a nanofluid of 5 vol. % concentration. (a) U configuration (b) I configuration (c) Z configuration.

Fig. 7 illustrates the temperature contours for U configuration at Re=100 and 5 kW/m$^2$ heat flux employing water and nanofluid as working fluids. From fig. 7 (a) it can be observed that for a U configuration under the influence of uniform thermal load, a hot zone creeps up at the intersection of the latter channels and the outlet manifold. The spatial spread of the hot zone is observed to be a direct function of the magnitude of heat flux and number of channels whereas inversely related to the flow Re and fluid thermal properties. In this particular configuration, the latter channels experience less mass flux of fluid and this leads to inefficient cooling towards the end of the latter channels. Increasing Re is a method to suppress such hot zones (which can over time deteriorate the life of the microdevice); however, it is not a feasible solution in practical applications since increased flow velocity might hamper the structural integrity of the device, often fabricated on semiconductor wafers. Accordingly, employing a working fluid with enhanced thermal properties is a solution and this is where nanofluids come into the forefront. As evident from fig. 7 (b), use of nanofluid not only suppresses the extent of the hot zone but also decreases the maximum temperature in the device. A reduction of ~ 4 °C at the core of the hot zone is observed when alumina based nanofluid is used instead of water. However, this too is not a complete picture of the capabilities of nanofluids. As reported by the present authors [17], DPM provides a more fundamentally accurate picture of the thermal performance of nanofluids within microscale flow domains. Hence, it is necessary to predict hot zones based upon DPM in cases where nanofluids are used to cool micro devices. It can be observed from Fig. 7 (c), that DPM predicts a further reduction in the size of hot zone and the maximum temperature is reduced by another ~ 2 °C compared to the EPM predictions.

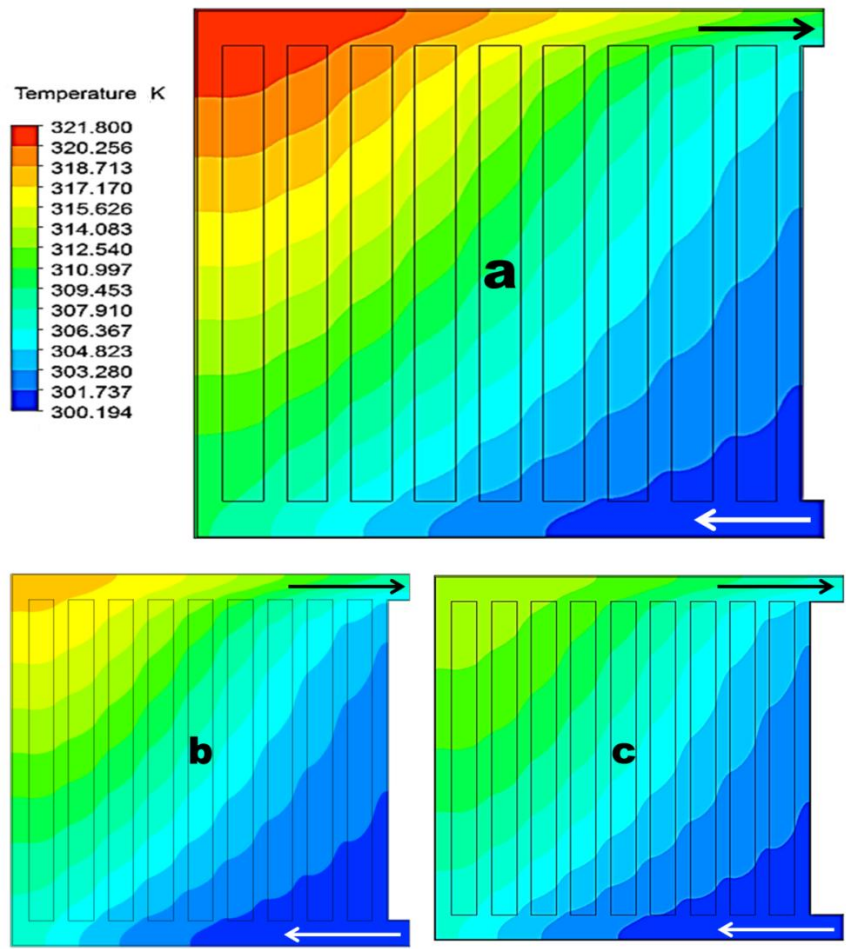

**FIGURE 7:** Thermal contours for U configurationat Re=100 and 5 kW/m$^2$ heat fluxwith working fluid as water and aqueous alumina (45 nm) nanofluids. (a) Water (b) nanofluid EPM and (c) nanofluid DPM.

Given the fact that the magnitude of temperature in the hot zone for U configuration, as well as the spread of the zone,are large, different flow domains can also be considered to counter the same. Fig. 8 illustrates the temperature contours for I configuration employing water and alumina–water at Re=100. From the figure it can be observed that there is formation of two hot zones but of relatively less temperature compared to that of U configuration. Since the flow within the manifold under goes bidirectional splitting, the net effect is somewhat similar to that of a U type configuration, but cut open into two. Consequently, in this case too, the end channels are devoid of requisite fluid flow and hence two hot zones symmetrically erupt at the two corners of the outlet manifold. However, it is noteworthy that the flow distribution in I configuration is better than that of U and thus the core temperature of the hot zones are appreciably lesser than U case. Similar to that discussed above, employing nanofluid as the working fluid will lead to lesser spread as well as reducedcore temperature of hot zone when compared to water. While the EPM approach leads to a reduction of ~ 2$^{o}$C with respect to water, the DPM predicts a further reduction of ~ 1.5 $^{o}$C. The relative reduction in the core temperature is lesser in case of I than that of U and this seemingly anomalous behavior can be explained on the basis of slip mechanisms of the nanoparticles. Since the hot zone temperatures in case of U are higher than that of I, the relative strength of the thermophoretic and Brownian diffusivities are higher in the former than the latter. This enhanced diffusivity leads to enhanced transport of heat from the hot walls to the comparatively colder bulk fluid and this leads the DPM approach to predict a relatively larger reduction in hot zone temperature in case of U than that of I.

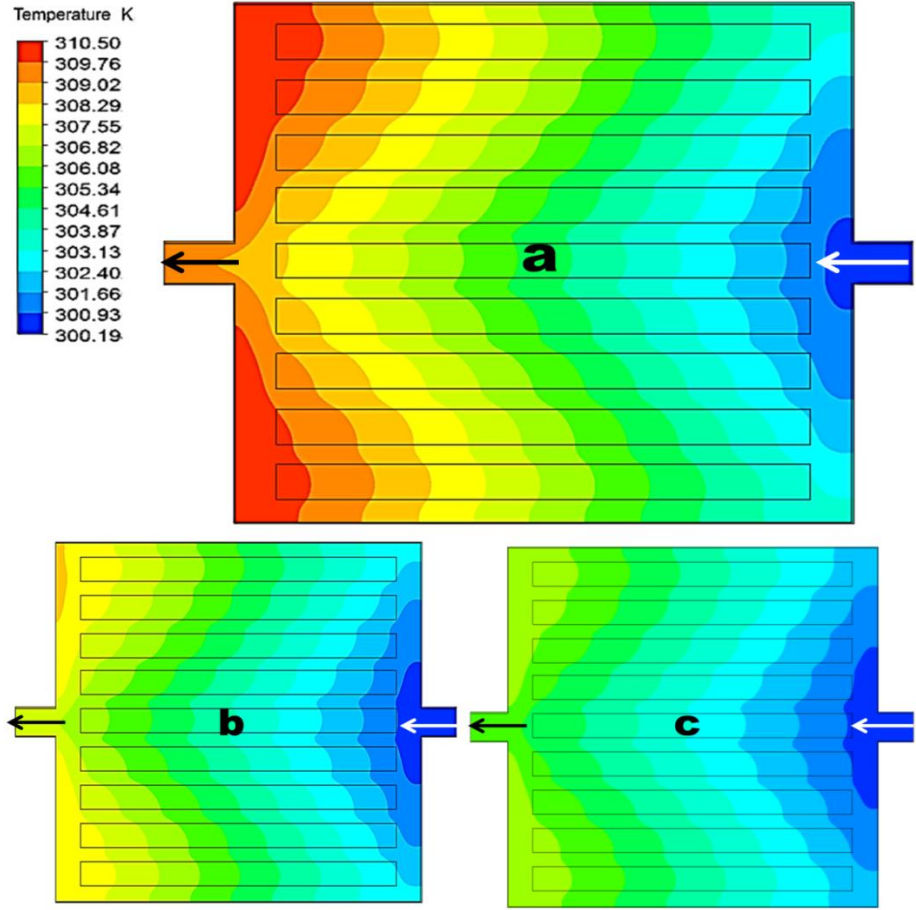

**FIGURE 8:** Thermal contours for I configuration at Re=100 and 5 kW/m$^2$ heat flux with working fluid as water and aqueous alumina (45 nm) nanofluid. (a) Water (b) nanofluid EPM and (c) nanofluid DPM.

The temperature contours of Z configuration for Re=100 and heat flux 5kW/m$^2$ have been illustrated in fig. 9. From the figure it can be observed that the location of hot zone is similar to the U configuration but with reduced hot zone core temperature. However, the extent of spreading of the hot zone is larger compared to U due to the differences in flow patterns in the outlet manifold. In U case, the direction of flow in the outlet manifold reverses with respect to the inlet manifold and hence the hot zone remains confined to the end of the outlet manifold. But in case of Z, the heat carried by fluid from the former channels convects into the hot zone, thereby elongating it along the outlet manifold. Nevertheless, it should also be noted, Z exhibits best flow distribution among the three and hence the mixing of relatively cold fluid from the former channels also reduces the temperature of the hot zone drastically. It can be observed from the figures that for a given thermal load and coolant mass flux and same design parameters, Z configuration provides best thermal management of the device. As it is familiar from earlier discussions, nanofluid will cool with higher efficacy than water. From fig. 9 (b) it can be observed that the maximum temperature of the hot zone reduced by ~1.5$^o$C and further by ~0.8$^0$C using DPM. This observation is consistent to that of I case since in case of Z the temperatures are further lower and hence the effect of DPM is reduced.

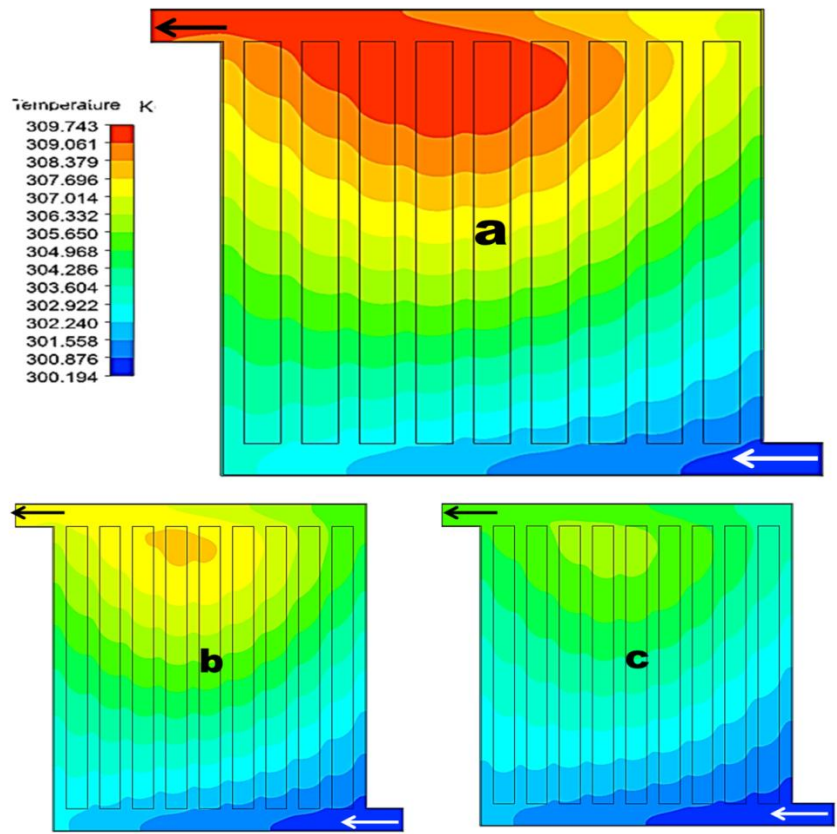

**FIGURE 9:** Thermal contours for Z configuration at Re=100 and 5 kW/m$^2$ heat flux with working fluid as water and aqueous alumina (45 nm) nanofluid. (a) Water (b) Nanofluid EPM and (c) Nanofluid DPM.

### 3.3. Thermal '*smart*' effects

The importance of Brownian and thermophoretic diffusions has been discussed in the preceding sections as well as reported by the present authors [17]. It can be further demonstrated that the cooling caliber of nanofluids increases in regions at and around the hot zones, thereby exhibiting potentially '*smart*' effects. The same has been illustrated in fig. 10 which has been plotted for I configuration at Re=100 and 12 kW/m$^2$ heat loademploying alumina-water (5 vol. %) as working fluid using DPM and EPM. Two horizontal sections A–A' and B–B' have been considered in such a way that the former one lies in hot region and the later one in a relatively cold region.Temperatures along the two sections have been plotted for DPM and EPM and have been superimposed onto the thermal contour for qualitative as well as quantitative insight. As observable, the difference between the temperatures predicted by DPM and EPM at a particular section is smaller in case of the plane B–B' than that of A–A' (shows by the arrows). A difference of ~ 0.7 – 0.8$^o$C can be observed between the mismatches of EPM and DPM between the two sections. Expectedly, the higher temperatures near the hot zone induces higher Brownian and thermophoretic drift fluxes onto the particles, enabling them to transport more heat than predictedby EPM. This essentially sheds insight onto the '*smart*' temperature response of nanofluids in microscale (since the drift flux would be negligibly minute in case of macroscale flows due to large characteristic flow length scales) flows.

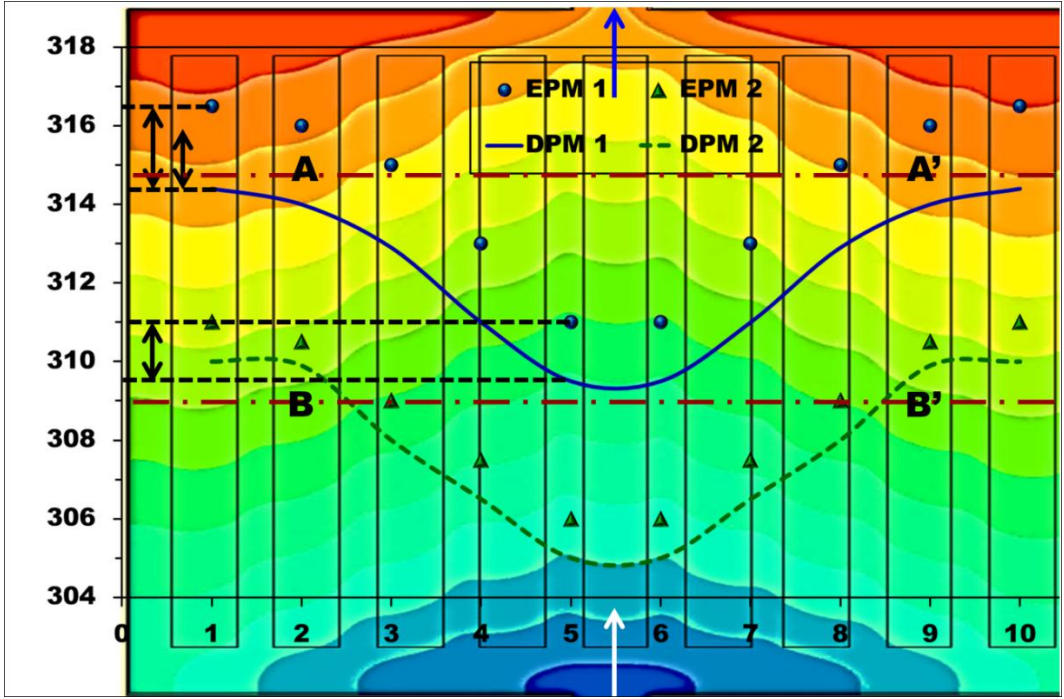

**FIGURE10:** Thermal performance of nanofluid using DPM and EPM.

Apart from the enhanced thermal performance due to usage of a nanofluid, it is also essential to understand the cooling capability of a system of microchannels solely by virtue of its configuration. It then becomes easy to shed insight onto the exact caliber of nanofluids as coolants in complex flow systems. Figure 11 illustrates the thermal contours for U, I and Z flow configurations for same thermal load and coolant flow conditions. In order to provide a qualitative comparison, the temperature scale has been presented equal for all cases with the maximum limit equal to the maximum hot zone temperature of U case. It can be observed that the Z flow pattern ensures best thermal management within the domain. The effect of thermophoretic drift and its implication vis-à-vis heat transfer can also be qualitatively seen when the concentration contours in the channels near or at the hot zones are considered. A clear effect of particle migration and consequent localized crowding can be observed at the exit section (it coincides perfectly with the hot zone) of the last but one channel in U case. However, similar phenomena is observed in the exit region of an extreme end channel in I since the hot zones are located therein. But it is noteworthy that since the hot zone temeprature in I is much reduced as compared to U, the crowding behavior is also qualitatively reduced. In case of Z, the crowding is more at the exit of last but one channel than the last channel because of the staggered hot zone as discussed earlier. However, here too, the crowding pattern is not as pronounced as compared to U.

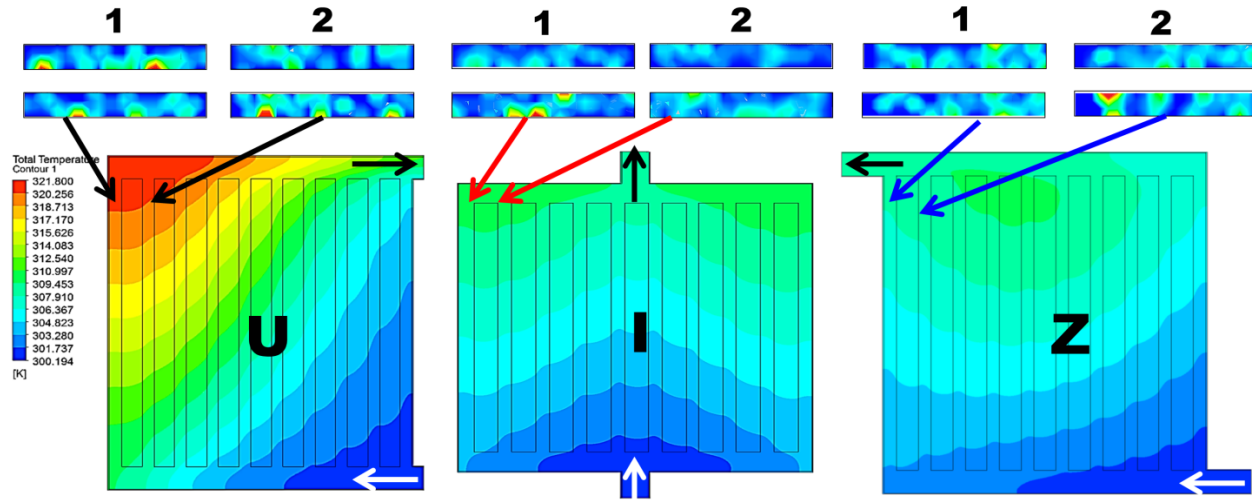

**FIGURE11:** Comparison of the thermal contours of the device for U, I and Z configuration microchannel systems at Re=100 and 5 kW/m$^2$ heat flux with alumina-water as working fluid. The efficacy of the flow configuration on thermal management capability has been illustrated. The effect of thermophoretic drift has been qualitatively shown using particle concentration contours at the corresponding hot zones. Effect of thermophoresis is clearly visible in the diabatic contours (lower row) when compared to the adiabatic contours (upper row) at the same zone. The arrows indicate the position of the cross section for which the contours have been shown.

The thermal management efficiency of a particular flow domain can be known from knowledge of mean temperature in the domain and the standard deviation of a statistically large population of temperature points in the domain. The results, as obtained from DPM and EPM have been illustrated in fig. 12. From the fig. 12 (a), it can be observed that the mean temperature for U configuration using water as well as nanofluid ismaximum when compared to Z and Iconfigurations, whichis expectedsince flow in U type is highly maldistributed. As the flow distribution improves in case of I and Z, the mean temperature within the device reduces significantly, and this can be further improved by the use of nanofluids.It is also noteworthy that although I and Z, by virtue of their configurations are more suited for cooling, the nonhomogeneous characteristic of nanofluids is better exploited in the U case. Due to the higher mean temperature in U, the magnitude of the slip forces (thermophoresis and Brownian) are large and the enhanced diffusivity leads to higher heat transfer augmentation by DPM than that predicted by EPM. It is observed that the drop in maximum temperature within the device when nanofluid (DPM) is employed instead of water is highest for the U configuration.

From an application point of view, uniformity in cooling performance is as important as magnitude of cooling. It can be observed from the figure that the Z configuration exhibits lowest mean temperature as well as standard deviation of the temperatures at a multitude of physical points within the device. It can thus be inferred that for systems generating heat loads in a more or less uniform manner, the Z flow configuration provides the best thermal management. Furthermore, a *Figure of Merit*, represented as $\xi = \frac{T_{max}}{T_{max}-T_{min}}$ has been defined to have more insight into the smart effect of nanofluids and capability of cooling uniformly. The figure of merit requires the temperatures to be expressed in the absolute scale. Figure 12 (b) represents the Figure of Merit for three different configurations. From the figure it is inferred that figure of merit of Z configuration is high in case of water as working fluid, followed by I and least for U configuration. Figure of merit further increases by using nanofluid as working fluid. Essentially, the parameter provides information on the cooling caliber of the system as well as the coolant and higher values imply higher caliber systems.

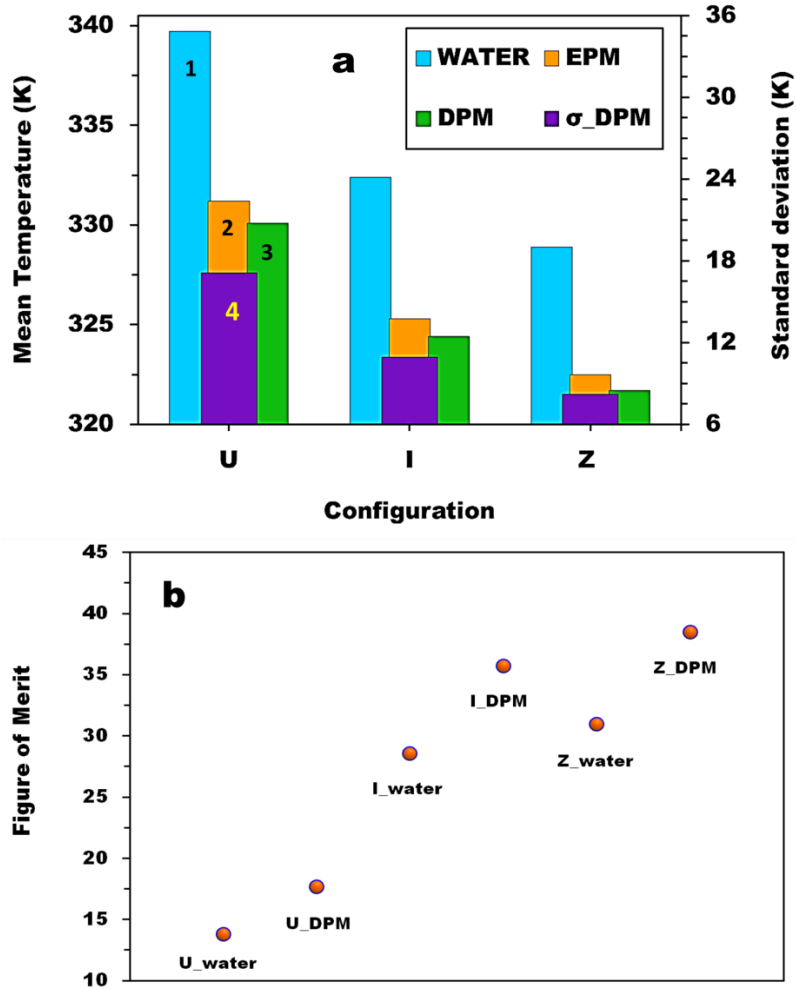

**FIGURE 12**: (a) Mean temperature of U, I and Z configurations with water and alumina-water nanofluid (5 vol. %) at Re=5 using DPM and EPM and standard deviation of temperature distribution (indicating uniformity in cooling) for DPM. (1, 2, 3 and 4 represents water, EPM, DPM, σ $_{DPM}$ respectively) (b) Figure of merit of three different concentrations using water and alumina (5 vol. %) –water nanofluid as working fluids at Re=100.

### 3.4. Non–uniform thermal dissipation

A simplistic thermal load model for the microelectronic device under consideration has been discussed until this point. In this case, the heat flux is considered to be spatially uniform, however, it is also necessaryto understand situations mimicking real life scenario where the thermal load is rarely uniform. In the present exercise, a silicon micro device has been considered with non-uniform heat fluxes and thermal load inputs closely mimic real time data of Intel® Core™ i7–4770(3.40 GHz) processor under various processor load conditions. The data has been extracted using a commercial software which provides real time mapping of the instantaneous power input and temperatures in 4 zones of the processor (for constant performance load of ~ 70% of total processor capacity) as well as displays the maximum power and temperature among the four zones for a given period of time. Figure 13 illustrates the thermal contours of U configuration with water as working fluid at Re=300 and under the influence of non–uniform heat load. As discussed in previous sections, the hot zone for U configuration with uniform heat flux can be expected to occur at the intersection of latter channels and the outlet manifold but the hot zone morphology may change in accordance to thermo fluidic parameters. However, this is not true in case of real situations where a particular zone of the chip might release higher heat due to higher performance of the electronic components in that region.

For the scenario where the high performance zone lies in position 1, 1 (refer fig. 13), a hot spot is seen to appear within the central region of the former channels. But predictions based on uniform heat flux show that the hot zone is only confined to the dead end of the outlet manifold and clearly it is not the case in the present scenario. This thereby justifies the need to study real time non–uniform heat flux cases for proper thermal design and management of microchips. As shown in fig. 13 (a), for active zone in 1,1 position, the channels are in general flushed by the coolant and hence despite the high heating (25 W), the hot zone temperature increases only by ~12 $^{o}$C as compared to uniform heating of 5W. When this zone of high heat dissipation shifts to position 1,2 (fig. 13 (b)), a drastic change in the temperature distribution pattern is observed. The core temperature now shoots up ~ 41$^{o}$C compared to the base performance condition and the hot zone spreads to cover almost a third of the chip area. As the position 1,2 encompasses the latter channels and U configuration experiences low fluid mass flux in the latter channels, the temperature not only shoots up in the active region but the heat is also convected along the latter channels such that the fluid can no further gather heat from the already fluid deprived 2,2 position. Hence, the hot zone spreads and covers a huge portion of the chip. In this scenario, the total area prone to thermal failure is the highest compared to all other scenarios considered.

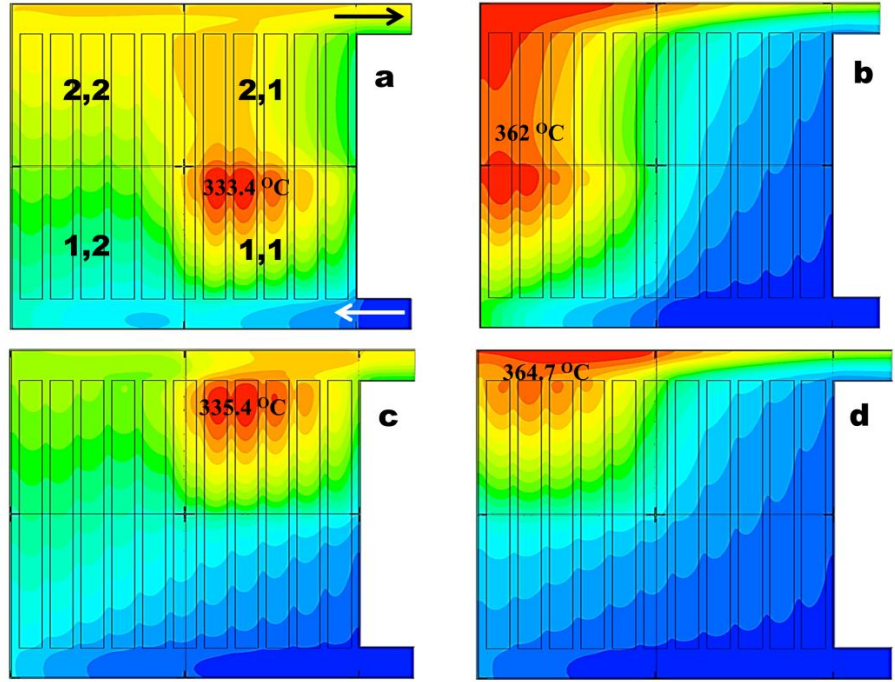

**FIGURE 13:** Illustration of thermal contours for non-uniform heat fluxes acting at different locations and its effect vis-à-vis core temperature, location and size of the hot zone. Water is used as the working fluid and the heat loads are 5W (for three zones on the device performing normal activity) and 25W (for the concerned higher activity zone). The location of the high activity zone is at (a) position 1, 1 (indicates $i^{th}$ heater of $j^{th}$ row from inlet) (b) position 1,2 (c) position 2,1 and (d) position 2, 2.

Figure 13 (c) illustrates the thermal contours corresponding to the case where 2, 1 is high heat dissipation zone. From the figure, it can be observed that the hot zone has shifted to the end of the former channels with some portion even encroaching into regions of the outlet manifold. In case of U configuration since fluid mass flux is more in the former channels, only a slight rise in temperatureis seen to occur in hot zone region, i.e. ~ 14°C when compared with reference case (uniform heat flux case). When the position of the high heat flux zone is shifted to the position 2, 2 as shown in fig. 13 (d), a rise in temperatureof ~ 43.7°C with respect to uniform heat flux caseis observed. It is due to the basic flow maldistribution pattern of U configuration. In this condition, the hot zone temperature is maximum, however, the regions of high temperature is restricted to a smaller zone than the case of 1,2 and so the probability of extensive thermal damage is lesser than 1,2 case.So it can be concluded that U configuration cannot be recommended for cases where high heat flux active zonesare in positions 1, 2 and 2, 2 (i.e. if it coincideswith the latter channels of device). The U configuration is suggested for non–uniform heat flux real scenario cases only where high heat dissipation zones are coincident with the former channels.Nanofluids are potential high caliber coolants and can be used to mitigate hot zone problems and fig. 14 illustrates the thermal contours using both water and nanofluid as coolants at Re=300. From the figure it can be observed that the use of nanofluid reduces the hot zone temperature as well as the size of hot zone considerably when compared with water. The enhanced thermophysical properties of the nanofluid as well as the particle migration mechanisms are responsible for higher diffusion of heat from the high temperature core and consequently the hot zone also shrinks in size.

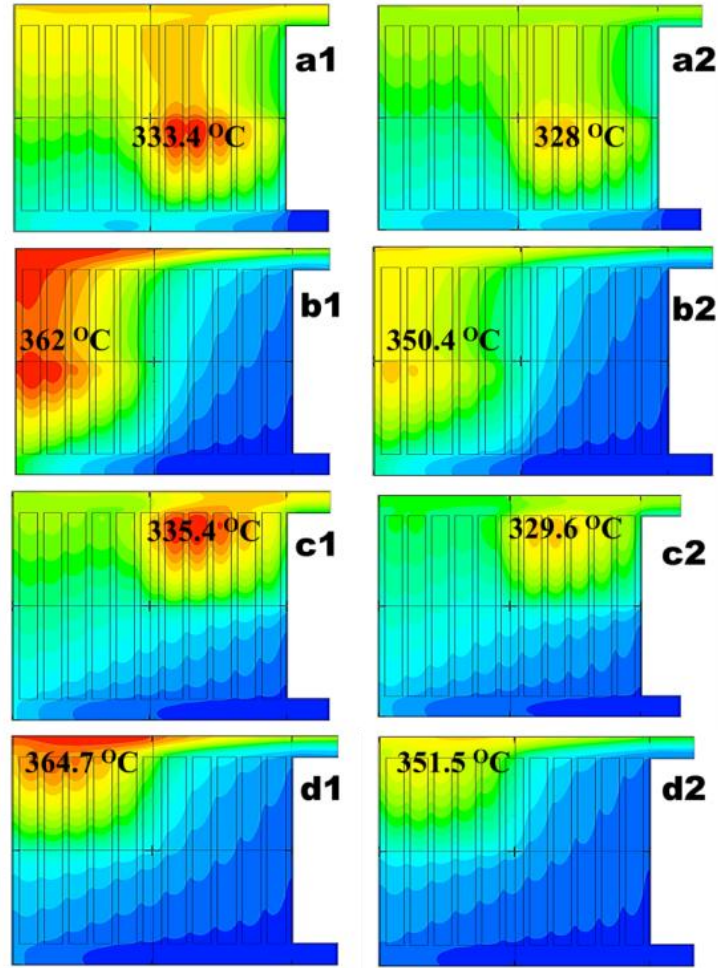

**FIGURE 14:** Illustration of the caliber of alumina-water (5 vol. %) nanofluids (DPM) and mitigating hot zones for non-uniform heat fluxes acting at different locations. Comparisons have been provided for the hot zone characteristics, both with respect to core temperature and morphology, using water (represented by 1) and nanofluids (represented by 2). The thermal conditions are same as that in Fig 13. The location of the high activity zone is at (a) position 1,1 (indicates $i^{th}$ heater of $j^{th}$ row from inlet) (b) position 1,2 (c) position 2,1 and (d) position 2, 2.

## 4. Conclusions

The present study focuses on the effects of flow and nanoparticle maldistribution characteristics of nanofluids (non-homogeneous formulation) and base fluids on the thermal performance of three different parallel microchannel cooling system configurations, i.e. U, I and Z. Literature available on nanofluid flow in microchannels model nanofluids as homogeneous fluids and consider effective properties to study enhanced heat transfer characteristics of the same in such complex devices. As reported by present authors [17], particle migration effects such as Brownian motion, thermophoresis, drag and gradients of pressure or shear stress are the major causes of non-homogeneous behavior of nanofluid in such confined microscale flow regions. Therefore, Discrete Phase Modeling has been employed to understand the behavior of nanofluid in three general configurations U, I and Z. The flow distribution in case of Z configuration is observed to be the most uniform followed by I and worst distribution has been observed for U configuration. Furthermore, the location of flow maldistribution generated hotspot with respect to flow configuration has been explored and it has been inferred that hotspot location and core temperature magnitude is a strong function of flow configuration in case of uniform heat flux cases. A *Figure of Merit* has been proposed to estimate the caliber of working fluid to effectively cool the given heat load uniformly. A high value of *Figure of Merit* is observed for Z configuration and further increase of value observed with nanofluid as coolant in the same configuration, proving conclusively that Z configuration is the most suitable one for such uniform thermal load device cooling. For mimicking a more real life scenario i.e. non uniform heat load over and above the maldistribution generated hot spots, data has been taken from an Intel® Core™ i7–4770(3.40 GHz) processor under various processor load conditions and have been applied to the worst flow distribution case (U configuration). It has been observed that nanofluids not only reduce the core temperature of the hot zones but also are instrumental in suppressing the size and extent of the hotspot. Nanofluids are conclusively observed to be efficient fluids for MEMS cooling via microchannel systems not only in the point of just removing heat but also in caliber to cool uniformly because of the temperature dependent '*smart*' nature.

## Acknowledgements

The authors thank the Defense Research and Development Organization (DRDO) of India for partial financial support for the computational facilities. LSM would like to thank the Ministry of Human Resource Development (Govt. of India) for the doctoral scholarship and IIT Ropar for facilitating partial support to the research work.